\documentclass[aps,twocolumn,superscriptaddress,altaffilletter,lengthcheck,
tightenlines,showpacs,showkeys]{revtex4}
\usepackage{amsmath} 
\usepackage{amssymb} 

\newcommand{\bt}{\beta}

\newcommand{\ben}{\begin{eqnarray}}
\newcommand{\een}{\end{eqnarray}}
\newcommand{\be}{\begin{equation}}
\newcommand{\ee}{\end{equation}}
\newcommand{\ba}{\begin{eqnarray}}
\newcommand{\ea}{\end{eqnarray}}
\newcommand{\n}{\label}

\newcommand{\ga}{\gamma}
\newcommand{\ro}{\rho}

\newcommand{\Om}{\Omega}

\newcommand{\bn}{\begin{equation}\label}

\usepackage[dvipdf]{epsfig}
\usepackage{color}

\begin{document}

\title{Nonbaryonic dark matter and scalar field coupled with a transversal interaction plus decoupled radiation}

\author{Luis P. Chimento}\email{chimento@df.uba.ar}
\affiliation{Departamento de F\'{\i}sica, Facultad de Ciencias Exactas y Naturales,  Universidad de Buenos Aires and IFIBA, CONICET, Ciudad Universitaria, Pabell\'on I, Buenos Aires 1428 , Argentina}
\author{Mart\'{\i}n G. Richarte}\email{martin@df.uba.ar}
\affiliation{Departamento de F\'{\i}sica, Facultad de Ciencias Exactas y Naturales,  Universidad de Buenos Aires and IFIBA, CONICET, Ciudad Universitaria, Pabell\'on I, Buenos Aires 1428 , Argentina}


\date{\today}
\bibliographystyle{plain}

\begin{abstract}

We analyze a universe filled with interacting dark matter, a scalar field accommodated as dark radiation along with  dark energy plus a decoupled radiation term within the framework of  spatially flat Friedmann-Robertson-Walker (FRW) spacetime. We work in a three-dimensional internal space spanned by the interaction vector  and use a transversal interaction  $\mathbf{Q_t}$ for solving the source equation in order  to find all the interacting component energy densities. We  asymptotically reconstruct the scalar field and potential from  an early radiation  era to the late dominate dark energy one,  passing through an intermediate  epoch dominated by dark matter. We apply the $\chi^{2}$ method to the updated observational Hubble data for constraining the cosmic parameters,  contrast with the Union 2 sample of supernovae, and analyze the amount of dark energy in the radiation era. It turns out that our model fulfills the severe bound of $\Omega_{\rm \phi}(z\simeq 1100)<0.018$ at $2\sigma$ level,  is consistent with the recent analysis that includes cosmic microwave background anisotropy measurements from the Atacama Cosmology Telescope and the South Pole Telescope along with   the future constraints achievable by Planck and CMBPol experiments, and  satisfies the stringent bound  $\Omega_{\rm \phi}(z\simeq 10^{10})<0.04$ at $2\sigma$ level  in the big-bang nucleosynthesis epoch.

\end{abstract} 
\vskip 1cm


\bibliographystyle{plain}

\maketitle



\section{Introduction}
Accurate observational tests confirm that the universe is currently  speeding up due to a gravitationally repulsive unknown agent called dark energy and it contributes nearly $70\%$ of the total energy of the universe \cite{Book}. The first evidence in favor of dark energy stems from measurement of brightness-redshift relation of supernovae (SNe) type Ia pointing out these exploding stars can be used as standard candles  for tracing a big portion of the cosmic evolution of the universe \cite{obse1}.  Cosmic microwave background (CMB) anisotropies \cite{obse2} are another observational data that have provided strong indirect  evidence in favor of  dark energy along with the fact that the spatial geometry of the universe is very close to being flat \cite{obse2}. At  present, there is a growing number of observational methods for probing the dynamical behavior of dark energy at different scales, for example, galaxy redshift surveys allow to obtain the Hubble expansion history by measurement of  baryon acoustic oscillation (BAO) in the galaxy distribution \cite{Book}, the geometric  weak lensing method applied to Hubble space telescope images helps to find tighter constraints on the  dark energy equation of state  also \cite{WL}. A second key  element in modern cosmology is dark matter, its origin is simple and comes from the need to understand the observed mismatch between gravitational mass and the visible mass of galaxies or clusters of galaxies \cite{Book}, \cite{DMobserva}, \cite{dme}. Since  galaxy clusters  are the largest structures in the universe that have undergone gravitational collapse these are considered as a remarkably useful window for testing the  content matter  of the universe; then the main astrophysical evidences  for dark matter come form colliding galaxies, gravitational lensing of mass distribution or power spectrum  of clustered matter \cite{dme}. The aforesaid astrophysical observations also suggest that dark matter is a substantial non-baryonic (invisible) component representing  nearly $25\%$ of the total energy-matter of the universe \cite{dme} and the major agent responsible for the large-structure formation in the universe \cite{Book}, \cite{DMobserva}. Even though we found the incredible role that dark matter has played for resolving the riddle of missing  mass since its discovery by Zwicky in 1933 \cite{dmo},  one open issue is a microscopic theory for describing dark matter, actually  many theoretical physics have claimed that dark matter could be probably a new heavy  weakly interacting particle unable to be detected by  particle accelerator or cosmic rays experiments\cite{DMobserva}. 

A natural question is whether  dark matter and dark energy can exchange energy between them, and if they do,  if such transfer of energy could alter the cosmic history,   leaving testable imprints in the universe \cite{jefe1}.  It is believed  that a coupling between dark energy and dark matter changes the background evolution of the dark sector allowing to constrain any type of interaction and giving  rise to a richer cosmological dynamics compared with non-interacting models \cite{jefe1}. A step forward for constraining dark matter and dark energy  is to use  the physic behind  recombination or big-bang nucleosynthesis epochs by adding a decoupled radiation term to the dark sector for taking into account the stringent bounds  related to  the behavior of dark energy at early times \cite{hmi1}, \cite{hmi2}. Another appealing possible way to extend the insight about the dark matter-dark energy interacting mechanism is to explore a bigger picture in which a third component is added \cite{T1}, \cite{T2},\cite{T3a},\cite{T3b}, \cite{T4}, \cite{T5}. 

Recently,  the authors have explored the behavior of a cosmological scenario in which a  spatially flat FRW universe contains  three interacting components, i.e., dark energy, dark matter, and dark radiation \cite{jefe2} plus a decoupled radiation term. Here, we follow Ref. \cite{jefe2} and develop a model with a transversal interaction  proportional to a linear combination of the  total energy density and its derivative up to third order. 
We perform a cosmological constraint using the updated Hubble data \cite{H29} of 29 points, the severe bounds for  dark energy at early times  \cite{EDE1}, \cite{Amendola} in the recombination era or nucleosynthesis epoch \cite{Cyburt}, \cite{Cyburt2},  compare the theoretical distance modulus $\mu(z)$  with   the Union 2 compilation   of supernovae Ia \cite{amanu} but we also compare  our  constraints on cosmological parameters with the bounds reported by Planck  measurements \cite{Planck2013} and WMAP-9 project \cite{WMAP9} . We will use units such that $8\pi G=1$ and signature for the metric of the spacetime is $(-,+,+,+)$.



\section{The model }

We consider a spatially flat homogeneous  and isotropic universe described by FRW spacetime with line element given by  $ds^{2}=-dt^{2}+a^{2}(t) (dx^{2}+dy^{2}+dz^{2})$ being  $a(t)$ the scale factor. The universe is  filled with a matter component  interacting with a scalar field, they have energy densities $\ro_m$ and $\ro_\phi$, respectively, plus a decoupled component $\ro_r$, so that the evolution of the FRW universe is governed by  the Friedmann and conservation equations,   
\be
\n{01}
3H^{2}=\ro_{\rm t}=\ro_{\rm m}+\frac{1}{2}\dot\phi^2+V(\phi)+\ro_{\rm r},
\ee
\be
\n{02}
\dot{\ro}+3H(\ro_{\rm m}+\dot\phi^2)=0,
\ee
\bn{rr}
\dot\ro_{\rm r}+3H\ga_{\rm r}\ro_{\rm r}=0,
\ee
where $H = \dot a/a$ is the Hubble expansion rate and 
\bn{ro}
\ro=\ro_{\rm m}+\frac{1}{2}\dot\phi^2+V(\phi),
\ee
includes all dark components, so $\ro_{\rm t}=\ro+\ro_{\rm r}$. Because of the additivity of the stress-energy tensor we identify the scalar field with a fluid having  energy density and pressure
\be
\n{rq}
\rho_\phi=\frac{1}{2}\dot\phi^2+V(\phi),
\qquad p_\phi=\frac{1}{2}\dot\phi^2-V(\phi),
\ee 
and describe it as a mix of two fluids, namely, $\rho_{\rm sm}=\dot\phi^2/2$ and $\rho_{\rm v}=V(\phi)$, with equations of state $p_{\rm sm}=\rho_{\rm sm}$ and $p_{\rm v}=-\rho_{\rm v}$ respectively. We assume that the equations of state have the barotropic form $p_{i}=(\gamma_{i}-1)\ro_{i}$, where the constants $\ga_i$ indicate the barotropic index of each component being $i={\rm v,m,sm,r}$, so   $\gamma_{\rm v}=0$, $\ga_{\rm m}=1$, $\gamma_{\rm sm}=2$ and $\ga_{\rm r}$ will be estimated later on. Then, $\rho_{\rm v}$ plays the role of some kind of vacuum energy, $\ro_{\rm m}$ represents a pressureless matter component, while $\rho_{\rm sm}$ can be associated with  stiff matter. 

Even in the absent of the matter component, we observe from Eq. (\ref{02}) that the two fluids in which the scalar field was split are interacting. In fact, we have
\bn{cv}  
\ro_{\rm v}'=Q
\ee
\bn{cf}
\ro_{\rm sm}'+2\ro_{\rm sm}=-Q,
\ee
where we have introduced the variable $\eta = \ln(a/a_0)^{3}$, with $a_0$ the present value of the scale factor, $' \equiv d/d\eta$, and the interaction term $3H Q$, which
describes the energy transfer between these components, so we find that $Q=dV(\phi)/d\eta$ is an intrinsic interaction, between the potential $V(\phi)$ and the corresponding kinetic energy $1/2\dot\phi^2$, which leads to the ordinary Klein-Gordon equation
\be
\label{fg}
\ddot{\phi}+3H\dot{\phi}+  V'(\phi)=0,
\ee
after summing the last two equations.

When additionally the matter  interacts with the scalar field, we introduce the three interaction terms $3H Q_{\rm v}$, $3H Q_{\rm m}$, and $3H Q_{\rm sm}$, so that the conservation equation (\ref{02}) is split into three balance equations
\be
\n{04}
\ro_{\rm v}'  =  Q_{\rm v},
\ee
\be
\n{05}
\ro_{\rm m}' + \ro_{\rm m} = Q_{\rm m},
\ee
\be
\n{06}
\ro_{\rm sm}' + 2\ro_{\rm sm} = Q_{\rm sm},
\ee
where the interaction terms satisfy the condition
\be
\n{06b}
 Q_{\rm v}+  Q_{\rm m}+ Q_{\rm sm}=0,
\ee
to recover the whole conservation equation
\be
\n{03}
\ro'=-\ro_{\rm m}-2\ro_{\rm sm}.
\ee
 
In this scalar field model, we have a mix of interacting vacuum energy, ordinary matter and stiff matter, where the vacuum energy and stiff fluid are accommodated as a scalar field, its dynamic being governed by a modified Klein-Gordon (MKG) equation, easily obtained  from the first derivative of  $\ro_{\phi}$  along with the help  of  Eq.  (\ref{05}):
\be
\label{re2}
\ddot{\phi}+3H\dot{\phi}+  V'(\phi)=\frac{-3HQ_{\rm m}}{\dot{\phi}}.
\ee
Essentially, this is an interesting model to describe a universe which transits between a dominated   stiff matter era at early times and a de Sitter scenario at late times. However, our goal in this paper is to construct a more realistic cosmological model where the dynamic of the universe is controlled by an early radiation-dominated era. Next, the radiation becomes subdominant with the Universe entering the dark matter  era  and subsequently followed by a
final stage where  dark energy dominates at late times. This realistic model can be achieved by means of a coupling between the three components, as  can be noticed from Eqs. (\ref{04})-(\ref{06}). In fact, there we have assumed constant barotropic indeces for the three components $(\ga_{\rm v}, \ga_{\rm m}, \ga_{\rm sm})=(0,1,2)$, however, the exchange of energy between them alters their own characteristics because the interaction terms $Q_{i}$ modified those initial values of $(\ga_{\rm v}, \ga_{\rm m}, \ga_{\rm sm})$, so that the effective barotropic indeces become $(-Q_{\rm v}/\ro_{\rm v}, 1-Q_{\rm m}/\ro_{\rm m}, 2-Q_{\rm sm}/\ro_{\rm sm})$, according to  Eqs. (\ref{04})-(\ref{06}). This allows us to choose an adequate set of interaction terms $Q_{\rm i}$ to modify the initial barotropic indexes, in a sense that the three interacting components represent radiation,  dark matter and dark energy respectively. To this end we adopt a transversal interaction $\mathbf{Q_t}$ between the three components. 

Let us accommodate the interaction terms $Q_{\rm i}$ as the components of a vector $\mathbf{Q}=(Q_{\rm v}, Q_{\rm m}, Q_{\rm sm})$ which lives on the interaction plane defined by  Eq. (\ref{06b}). Thus, we introduce a three-dimensional internal space with an orthonormal vector basis $\{\mathbf{e_t},\mathbf{e_o},\mathbf{n}\}$ and set the coordinate origin at the intersection of the plane (\ref{06b}) with the barotropic index vector  $\mbox{\boldmath ${\gamma}$}=(\ga_{\rm v},\ga_{\rm m},\ga_{\rm sm})$. The pair of vectors $\{\mathbf{e_t},\mathbf{e_o}\}$ is contained in the interaction plane,  $\mathbf{e_t}$ is orthogonal to the vector $\mbox{\boldmath ${\gamma}$}$, the orthogonal projection of the vector $\mbox{\boldmath ${\gamma}$}$ on the interaction plane defines the direction of $\mathbf{e_o}$ and $\mathbf{n}$ is orthogonal to the interaction plane, meaning that $\mathbf{Q}=q_t\,\mathbf{e_t}+q_o\,\mathbf{e_o}$, where $q_t$ and $q_o$ are the components of the interaction vector $\mathbf{Q}$ on the plane (\ref{06b}).

With $\mathbf{e_t}$ the unique vector of the basis orthogonal to the vector $\mbox{\boldmath ${\gamma}$}$, we select only those interactions which are collinear with the aforesaid preferred direction in the plane  (\ref{06b}) 
$\mathbf{Q_t}=q_t\,\mathbf{e_t}$ which will be called ``transversal interaction'', ensuring that $\mbox{\boldmath ${\gamma}$}\cdot \mathbf{Q}=0$. In our case, the vector barotropic index is $\mbox{\boldmath ${\gamma}$}=(0,1,2)$ and the transversal interaction becomes  
\bn{qt}
\mathbf{Q_t}=(1,-2,1)Q_{\rm v}.
\ee
In the following we will see that the transversal character of the interaction vector (\ref{qt}) will simplify enough the algebra of the model for more details see Ref. \cite{jefe2}. For instance from the transversal interaction (\ref{qt}) we have $Q_{\rm sm}=Q_{\rm v}$, then inserting it in Eq. (\ref{06}) we obtain
\be
\n{smi}
\ro_{\rm sm}' + 2\ro_{\rm sm} = Q_{\rm v},
\ee 
Now, if we compare the scalar field  interacting  with an intrinsic interaction $Q$ in absence of dark matter, represented by  Eqs. (\ref{cv}) and (\ref{cf}), with the same scalar field in presence of an explicit transversal interaction between this component and dark matter see Eqs. (\ref{04}) and (\ref{smi}), then we conclude that  the potential is not modified by the transversal interaction because Eqs. (\ref{cv}) and (\ref{04}) have the same meaning: that $Q=Q_{\rm v}$ , however, the kinetic energy is strongly affected as can be seen from Eqs. (\ref{cf}) and (\ref{smi}). In fact, both equations are not the same unless we make the transformation $\ro_{\rm sm}\to -\ro_{\rm sm}$, in this case the kinetic term changes its sign and the scalar field becomes imaginary. Besides, from the equations of state of the kinetic
energy, $p_{\rm sm}=\rho_{\rm sm}$, we reach the conclusion that the
pressure simultaneously changes its sign $p_{\rm sm}\to -p_{\rm sm}$  as
a
 consequence of the interaction between the dark fluid components.
Interestingly,  the change of sign of the kinetic energy does not emerge
when the flat Friedmann universe crosses a soft singularity as  occurs
for the tachyon fields \cite{kami}
because both, the associated  pressure $p_{\rm sm}$ and the  kinetic
energy $\rho_{\rm sm}$ diverge simultaneously according to  $p_{\rm
sm}=\rho_{\rm sm}$. 
Actually this is not a serious difficulty because we will see below that
the linear  transversal  interaction makes the energy density of the
scalar field $\rho_\phi=\dot\phi^2/{2}+V(\phi)$ be positive definite.
An important consequence of this result also can be extracted from Eqs. (\ref{04}) and (\ref{qt}), after combining them we reach  $Q_{\rm m}=-2Q_{\rm v}=-2\ro_{\rm v}'=-2(dV/d\phi)\,(d\phi/d\eta)=-2(dV/d\phi)\,(d\phi/3Hdt)$. Thus, the MKG equation (\ref{re2}) becomes 
\be
\label{re2'}
\ddot{\phi}+3H\dot{\phi}- V'(\phi)=0,
\ee
showing that the transversal interaction changes the sign in the derivative of the potential of the ordinary KG equation. In other words we have found the explicit transversal interaction
\bn{int}
-\frac{3HQ_{\rm m}}{\dot\phi}=2\,\frac{dV}{d\phi},
\ee
which modifies the behavior of the scalar field. The idea of the gradient of the potential generating an interaction between the two  perfect fluids, as was  mentioned above,  is not new and it was analyzed   in the literature  a long time ago (see e.g. \cite{mimo} and reference therein).  In order to link our formalism of interacting three components with the standard proposal examined within  the framework of interacting dark sector we must identify the exchange of energy encoded in Eq. (\ref{int})  with $Q_{m}=-2\,dV\,d\phi (2\rho_{\rm sm}/3\rho_{\rm t})^{1/2}=[\partial \ln m(\phi)/\partial \phi] \rho_{a}\dot{\phi}$, where the choice of the mass function $ m(\phi)$ defines the coupling and as a consequence the interaction itself \cite{Amendola2}, and $\rho_{a}$ could be the total energy density or one of its components.  

After differentiation Eq. (\ref{03}) and using Eqs. (\ref{04})-(\ref{06}) and (\ref{qt}), we obtain   
\be
\n{07}
\ro''=\ro_{\rm m}+ 4\ro_{\rm sm}.
\ee
Then, we will construct an interacting three fluid model with the transversal interaction (\ref{qt}) by solving the algebraic system of equations in the $(\ro_{\rm v}, \ro_{\rm m}, \ro_{\rm sm})$-variables (\ref{ro}), (\ref{03}), and (\ref{07}) to find $\ro_{\rm v}$,  $\ro_{\rm m}$, and  $\ro_{\rm sm}$ as functions of $\ro$, $\ro'$ and $\ro''$, 
\be
\n{08}
\ro_{\rm v}=\frac{1}{2}\left[ 2\ro+3\ro'+ \ro''\right],
\ee
\be
\n{09}
\ro_{\rm m}=-\left[ 2\ro'+ \ro''\right],
\ee
\be
\n{10}
\ro_{\rm sm}=\frac{1}{2}\left[\ro'+\ro''\right],
\ee
Following  Refs. \cite{jefe1} and \cite{jefe2}, we replace (\ref{08}) into (\ref{04}), (\ref{09}) into (\ref{05}) or (\ref{10}) into (\ref{06}) and get the third order differential equation, which we call ``source equation'', for the total energy density: 
\be
\n{11}
\ro'''+3\ro''+2\ro'=2Q_{\rm v}.
\ee
Thus, once the transversal interaction $\mathbf{Q_t}$ is specified, we obtain the energy density $\ro$ by solving  the source equation  (\ref{11}), whereas the component energy densities $\ro_{\rm v}$, $\ro_{\rm m}$, and $\ro_{\rm sm}$ are obtained after inserting $\ro$, $\ro'$, and $\ro''$ into Eqs. (\ref{08})-(\ref{10}). 


\subsection{Linear  transversal  interaction $\mathbf{Q_t}$ }

We will look for the set of transversal interaction which are linearly dependent on $\ro_{\rm v}$, $\ro_{\rm m}$, $\ro_{\rm sm}$, along with their derivatives up to first order, and $\ro$, $\ro'$, $\ro''$, $\ro'''$. Hence, after using Eqs. (\ref{08})-(\ref{10}) one finds that 
\be
\label{qv}
Q_{\rm v}=\beta_{1}\ro+\bt_2\ro'+\beta_{3}\ro''+\beta_{4}\ro''',
\ee
becomes a linear functional of the basis elements  $\ro$, $\ro'$, $\ro''$, $\ro'''$ (see \cite{jefe1}, \cite{jefe2}), and $\beta_{1}$, $\bt_2$, $\beta_{3}$, $\beta_{4}$ are four constant parameters which will be restricted to the ranges where the characteristics roots of the source equation (\ref{11}) are positive definite. The energy density $\ro$ is found by solving the source equation (\ref{11}) for the general linear transversal interaction (\ref{qv}). It is given by 
\be
\n{DenZ}
\ro=3H^{2}_{0}\Big({\cal A}x^{3\ga_{s}}+{\cal B}x^{3\ga_{+}}+{\cal C}x^{3\ga_{-}}\Big),
\ee
where $x=z+1$ and  $z$ is the cosmological redshift while the characteristic roots $(\ga_s,\ga_-,\ga_+)$ of the source equation (\ref{11}), sourced by the interaction $Q_{\rm v}$ (\ref{qv}), are accommodated according to the requirement that $\ga_s$ is the minimum of $\{\ga_s,\ga_-,\ga_+\}$. The component energy densities become
$$
\ro_{\rm v}\to\ro_{\rm x}=\frac{3H^{2}_{0}}{2}[(\ga_s-2)(\ga_s-1){\cal A}x^{3\ga_{s}}
$$
\bn{vf}
+(\ga_+-2)(\ga_+-1){\cal B}x^{3\ga_{+}}+(\ga_--2)(\ga_--1){\cal C}x^{3\ga_{-}}],
\ee
$$
\ro_{\rm m}\to\ro_{\rm c}=3H^{2}_{0}[\ga_s(2-\ga_s){\cal A}x^{3\ga_{s}}
$$
\bn{vm}
+\ga_+(2-\ga_+){\cal B}x^{3\ga_{+}}+\ga_-(2-\ga_-){\cal C}x^{3\ga_{-}}],
\ee
$$
\ro_{\rm sm}\to\ro_{\rm dr}=\frac{3H^{2}_{0}}{2}[\ga_s(\ga_s-1){\cal A}x^{3\ga_{s}}
$$
\bn{vsm}
+\ga_+(\ga_+-1){\cal B}x^{3\ga_{+}}+\ga_-(\ga_{-}-1){\cal C}x^{3\ga_{-}}].
\ee
after insert the total energy density (\ref{DenZ}) into Eqs. (\ref{08})-(\ref{09})-(\ref{10}).  
We have made the substitution $\ro_{\rm v}\to\ro_{\rm x}$, $\ro_{\rm m}\to\ro_{\rm c}$, and  $\ro_{\rm sm}\to\ro_{\rm dr} $ to show that the interaction was taken into account explicitly. More precisely, $\ro_{\rm x}$, 
 $\ro_{\rm c}$, and $\ro_{\rm dr}$ will represent the dark energy, dark matter and dark radiation energy densities. So that, we will work under the condition $\ga_{s}<2/3<\ga_{-}\approx 1<\ga_{+}\approx 4/3$ to show the transition of the universe from an early era dominated by radiation to an intermediate stage dominated by dark matter (non-baryonic) to end in a dark energy dominated era at late times. At this point it is interesting to remark that in the exceptional case $\ga_s<0$, the scale factor could have phantom behavior with a Big Rip singularity in remote future because and the energy density  is increasing without limit, meaning that $\ro_{\rm t}$ diverges at the singularity. Although in our model there is no phantom behavior because the best-fit values  for  $\ga_{s}$ are always positive definite, when we take into account the values of $\ga_{s}$ at $1\sigma$ or $2\sigma$ levels [see Table (\ref{I})] there is some probability that the model leads to a phantom scenario, indicating that such kind of final fate cannot be statistically excluded  at all.   

On the other hand, the integration constants ${\cal A}$, ${\cal B}$ and ${\cal C}$ can be expressed in terms of the density parameters

\[{\cal A}=\frac{(1-\Omega_{\rm \phi 0})(\ga_{+}-1)(\ga_{-}-1)+\Omega_{\rm r0}(\ga_{+}-2)(\ga_{-}-2)}{(\ga_{s}-\ga_+)(\ga_{s}-\ga_{-})}\]
\be
\n{A}
+\frac{(\Omega_{\rm \phi 0}-\Omega_{\rm r0})\ga_{+}\ga_{-}}{(\ga_{s}-\ga_{+})(\ga_{s}-\ga_{-})}
\ee
\[{\cal B}=\frac{(1-\Omega_{\rm \phi 0})(\ga_{s}-1)(1-\ga_{-})+\Omega_{\rm r0}(\ga_{s}-2)(2-\ga_{-})}{(\ga_{s}-\ga_{+})(\ga_{+}-\ga_{-})}\]
\be
\n{B}
-\frac{(\Omega_{\rm \phi 0}-\Omega_{\rm r0})\ga_{s}\ga_{-}}{(\ga_{s}-\ga_{+})(\ga_{+}-\ga_{-})}
\ee
\[{\cal C}=\frac{(1-\Omega_{\rm \phi 0})(\ga_{s}-1)(1-\ga_{+})+\Omega_{\rm r0}(\ga_{s}-2)(2-\ga_{+})}{(\ga_{s}-\ga_{-})(\ga_{-}-\ga_{+})}\]
\be
\n{C}
-\frac{(\Omega_{\rm \phi 0}-\Omega_{\rm r0})\ga_{s}\ga_{+}}{(\ga_{s}-\ga_{-})(\ga_{-}-\ga_{+})}
\ee
where $\Omega_{\rm v0}=\ro_{\rm v0}/3H^{2}_{0}$, $\Omega_{\rm dr0}=\ro_{\rm dr0}/3H^{2}_{0}$, $\Omega_{\rm c 0}=\ro_{\rm c0}/3H^{2}_{0}$, and  $\Omega_{\rm r 0}=\ro_{\rm r0}/3H^{2}_{0}$  are the density parameters  fulfilling the condition $\Omega_{\rm \phi 0}+\Omega_{\rm c0}+ \Omega_{\rm r0} =1$ for a spatially flat FRW universe along with  $\Omega_{\rm v0}+\Omega_{\rm dr0}=\Omega_{\rm \phi 0}$. 

\subsection{Reconstructing the scalar field}

Due to the interest of reconstructing the scalar  field in the FRW cosmology we will implement a useful procedure to find the potential as a function of the scalar field $\phi$ by taking advantage of the knowledge of $V(a)$, $\dot\phi^2(a)$ and $\ro_t(a)$ as explicit function of the scale factor instead of the cosmological time through $V=\ro_{\rm x}$, $\dot\phi^2/2=\rho_{\rm dr}$ and $\ro_{\rm t}=\dot\phi^2/2+V+\ro_{\rm m}+\ro_{r0}/a^{3\ga_{\rm r}}$, we extract this information from Eqs. (\ref{vf})-(\ref{vsm}). Rewritten the kinetic energy density $\dot\phi^2/2$  of the scalar field as $\dot\phi^2=3\ro_t\phi'^2$, we easily obtain it as a function of the scale factor
\bn{re3}
\phi=\int{\frac{\sqrt{6\Om_{\rm sm}}}{a}\,\,da},
\ee
where $\Om_{\rm sm}=\dot\phi^2/2\ro_{\rm t}$. After integrate the last equations we find the scalar field $\phi(a)$. Inverting it gives $a(\phi)$ and by using  Eq. (\ref{vf}), it follows $V(\phi)=\ro_{\rm x}(a(\phi))$. Also,  we can reconstruct the scalar field and potential in terms of  redshift $z=-1+a^{-1}$. Thus the procedure determines $V(\phi)$ and defines a model with an exact solution for the spatially flat FRW cosmology. In addition, the energy density of the scalar field $\ro_\phi=V+\dot\phi^2/2=\ro_{\rm x}+\rho_{\rm dr}$ as a function of the scale factor is easily calculated from Eqs. (\ref{vf}) and (\ref{vsm}), it reads
$$
\ro_{\phi}=3H^{2}_{0}[(\ga_s-1)^2{\cal A}x^{3\ga_{s}}
$$
\bn{rphi}
+(\ga_+ -1)^2{\cal B}x^{3\ga_{+}}+(\ga_- -1)^2{\cal C}x^{3\ga_{-}}],
\ee
showing that $\ro_\phi$ is a real function in spite of the scalar field may take imaginary values. Resuming in our model, not only the total energy density of the scalar field $\ro_\phi=\dot\phi^2/2 +V=\ro_{\rm x}+\ro_{\rm dr}$ remains positive definite always, but most importantly, the  radiation and dark energy components are well defined along the stage we were considered. 

As a last comment, we have made an artificial splitting of the scalar field component into a potential $V(\phi)$ and a kinetic energy density $1/2 \dot{\phi}^2$ terms, which were associated with  the energy densities $\rho_{\rm v}$ and $\rho_{\rm sm}$, respectively. Basically in this association we have changed the original degree of freedom $\phi$ by the scale factor $a$ as can be seen from Eqs. (\ref{vf}), (\ref{vsm}) and (\ref{rphi}). The replacement of the degree of freedom was essential to facilitate the reconstruction process at the beginning of this section. This is a natural consequence of the interaction model that we have used, moreover, the modeling done in this section and the forthcoming one, where we will deal with the observational constraints, require an explicit scale factor dependence of all magnitudes and principally of the Hubble expansion rate.


Let us apply the reconstruction procedure to analyze the asymptotic behaviors of the scalar field and potential in the early radiation-dominated era, where the overall barotropic index $\ga\simeq \ga_+\approx 4/3$, and in the dark energy dominated era where the universe has an accelerated expansion implying  $\ga\simeq\ga_s$. At early times, from Eqs. (\ref{vf}) and (\ref{re3}) we have the approximated potential and scalar field 
\bn{pr}
V\simeq \frac{3H^{2}_{0}}{2}(\ga_+-2)(\ga_+-1)\,{\cal B}\,a^{-3\ga_{+}},
\ee
\bn{fr}
\Delta \phi \simeq -\sqrt{3 \ga_{+} (\ga_{+}-1)}\ln a,
\ee
the former being negative definite because the effective barotropic index $\ga\approx \ga_{+}$ is close to $4/3$ and the latter becomes real while the scale factor behaves as $a\approx t^{2/3\ga_+}$. Hence reconstructing the potential and the interaction term $Q_{\rm m}$, we obtain 
\bn{vrr}
V\simeq \frac{3H^{2}_{0}}{2}(\ga_+-2)(\ga_+-1)\,{\cal B}\,e^{\sqrt{3 \ga_{+}/ (\ga_{+}-1)}\,\Delta\phi},
\ee
and
\bn{qm}
Q_{\rm m}\simeq-3H^{2}_{0}\ga_+(\ga_+-2)(\ga_+-1)\,{\cal B}\,e^{\sqrt{3 \ga_{+}/ (\ga_{+}-1)}\,\Delta\phi},
\ee
after calculating the explicit transversal interaction (\ref{int}) in the early limit.

At late times, when the dark energy  governs the dynamic of the universe,  Eqs. (\ref{vf}) and (\ref{re3}) lead to 
\bn{ps}
V \simeq \frac{3H^{2}_{0}}{{2}}(\ga_s-2)(\ga_s-1)\,{\cal A}\,a^{-3\ga_{s}},
\ee
\bn{fs}
\Delta \phi\simeq\sqrt{3 \ga_{s}  (\ga_{s}-1 )}\ln a,
\ee
hence, after reconstruct the potential, we obtain the following positive real expression:  
\bn{psr}
V \simeq \frac{3H^{2}_{0}}{{2}}(\ga_s-2)(\ga_s-1)\,{\cal A}\, e^{ -\sqrt{3 \ga_{s}/ (\ga_{s}-1 )}\,\Delta\phi},
\ee
because  the effective barotropic index $\ga\approx \ga_s$. In this late regimen the scale factor behaves as $a\approx t^{2/3\ga_s}$. Using again the explicit transversal interaction (\ref{int}) in this late approximation we get the interaction term $Q_{\rm m}$
\bn{qm'}
Q_{\rm m}\simeq -3H^{2}_{0}\ga_s(\ga_s-2)(\ga_s-1)\,{\cal A}\,e^{\sqrt{3 \ga_{s}/ (\ga_{s}-1)}\,\Delta\phi},
\ee

In the intermediate stage, where effective barotropic index $\ga\simeq\ga_-\approx 1$ the universe transits a dark matter dominate era and the dark matter energy density behaves as $\ro_{\rm c} \approx 3H^{2}_{0}{\cal C}a^{-3}$ while  the scalar field and potential become negligible.  

\section{Observational constraints on a transversal interacting  model}
We will provide a qualitative estimation of the cosmological parameters by constraining them with the Hubble data  \cite{obs3}- \cite{obs4} and the strict bounds for the behavior of dark energy at early times \cite{EDE1},\cite{EDE2}, \cite{Amendola} . In the former case, the  statistical analysis is based on the $\chi^{2}$--function of the Hubble data which is constructed as (e.g.\cite{Press})
\be
\n{c1}
\chi^2(\theta) =\sum_{k=1}^{29}\frac{[H(\theta,z_k) - H_{\rm obs}(z_k)]^2}{\sigma(z_k)^2},
\ee
where $\theta$ stands for cosmological parameters, $H_{\rm obs}(z_k)$ is the observational $H(z)$ data at the redshift $z_k$, $\sigma(z_k)$ is the corresponding $1\sigma$ uncertainty, and the summation is over the $29~$ observational  $H(z)$ data\cite{H29}. The Hubble function is not integrated over and it is directly related with the properties of the dark energy, since its value comes from the cosmological observations. Using the absolute ages of passively evolving galaxies observed at different redshifts, one obtains the differential ages $dz/dt$ and the function $H(z)$ can be measured through the relation $H(z)=-(1+z)^{-1}dz/dt$ \cite{obs3}, \cite{obs4}. The data  $H_{\rm obs}(z_i)$ and $H_{\rm obs}(z_k)$ are uncorrelated because they were obtained from the observations of galaxies at different redshifts. 

From Eq. (\ref{DenZ}) one finds that the Hubble expansion of the model  becomes
\be
\n{Ht}
H(\theta| z)=H_{0} \Big( {\cal A}x^{3\ga_{s}}+{\cal B}x^{3\ga_{+}}+{\cal C}x^{3\ga_{-}}+ {\cal D}x^{3\ga_{r}}\Big)^{\frac{1}{2}}
\ee
${\cal A}$,  ${\cal B}$, and  ${\cal C}$ being obtained form (\ref{A}), (\ref{B}), and (\ref{C}),  respectively,  but  ${\cal D}$ is obtained from  (\ref{rr}). Here, we consider  $\theta=\{H_{0},\ga_{\rm s},  \ga_{+}, \ga_{-},\ga_{\rm r}, \Omega_{\rm  v0},\Omega_{\rm dr0}, \Omega_{\rm c0}\}$ as the independent parameters to be constrained  for the model encoded in the Hubble function (\ref{Ht}) with the statistical estimator (\ref{c1}). We will take  two independent parameters and fix the other ones along the statistic analysis until all parameters have been varied  and estimated with the $\chi^{2}$--function. Then, for a given pair of $(\theta_{1}, \theta_{2})$,  we are going to perform the statistic analysis by minimizing the $\chi^2$ function to  obtain the best-fit values of  the random variables $\theta_{\rm crit}=\{\theta_{\rm crit1}, \theta_{\rm crit2} \}$ that correspond to a minimum of Eq.(\ref{c1}). Then, the  best--fit parameters $\theta_{\rm crit}$ are those values where $\chi^2_{\rm min}(\theta_{\rm crit})$ leads to the local minimum of the $\chi^2(\theta)$ distribution. If $\chi^2_{\rm d.o.f}=\chi^2_{\rm min}(\theta_{\rm crit})/(N -n) \leq 1$ the fit is good and the data are consistent with the considered model $H(z;\theta)$. Here, $N$ is the number of data and $n$ is the number of parameters \cite{Press}. The variable $\chi^2$ is a random variable that depends on $N$ and its probability distribution is a $\chi^2$ distribution for $N-n$ degrees of freedom. Besides, $68.3\%$ confidence  contours  in the 2D plane  are made of the random data sets that satisfy the inequality $\Delta\chi^{2}=\chi^2(\theta)-\chi^{2}_{\rm min}(\theta_{\rm crit})\leq 2.30$. The latter equation defines a bounded region by a closed area around $\theta_{\rm crit}$ in the two-dimensional parameter plane, thus the $1\sigma$ error bar can be identified with the distance from the $\theta_{\rm crit}$ point to the boundary of the  two-dimensional parameter plane. It can be shown that $95.4\%$ confidence contours  with a $2\sigma$  error bar in the samples satisfy $\Delta\chi^{2}\leq 6.17$. Here $N=29$ and $n=8$, so in principle, we will perform   28 minimization of    $\chi^{2}$ statistical estimator, interpreting the goodness of  fit by checking the condition $\chi^2_{\rm d.o.f}<1$; as a way to keep the things clear and focus on extracting relevant physical information form this statistical estimation,  we only show the most interesting cases.

\begin{center}
\begin{table}
\begin{minipage}{1\linewidth}
\scalebox{0.45}{
\begin{tabular}{|l|l|l|l|l|l|l|l|l|l|l|}
\hline
\multicolumn{3}{|c|}{2D Confidence level} \\
\hline
Priors & Best fits $\pm 1\sigma \pm 2\sigma$ & $\chi^{2}_{\rm d.o.f}$ \\
\hline
{$(\Omega_{\rm v0}, \Omega_{\rm dr0},\ga_{s}, \ga_{+}, \ga_{-}, \ga_{\rm r})=(0.7499, 0.00006, 0.010, 1.3000, 1.0099, 1.3400)$}& ($H_{0}, \Omega_{\rm c0})= (70.00^{+3.76+4.82}_{-2.02-3.11}, 0.239^{+0.086+0.101}_{-0.019-0.042})$& $0.7368$\\
{$(\Omega_{\rm c0}, \Omega_{\rm dr0},\ga_{s}, \ga_{+}, \ga_{-}, \ga_{\rm r})=(0.1999,0.00006, 0.010, 1.3000, 1.0099, 1.3400)$}& ($H_{0}, \Omega_{\rm v0})= (70.00^{+4.99+6.07}_{-2.03-3.16}, 0.7299^{+0.057+0.0658}_{-0.0109-0.0238})$& $0.7368$\\
{($\Omega_{\rm v0}, \Omega_{\rm c0},\ga_{s}, \ga_{+}, \ga_{-}, \ga_{\rm r})=(0.7499, 0.1999, 0.010, 1.3000, 1.0099, 1.3400)$}& ($H_{0}, \Omega_{\rm dr 0})= (70.00^{+3.54+4.60}_{-2.00-3.06}, 0.00006^{+0.00666+0.14974}_{-0.02882-0.03404})$& $0.7368$\\
{($\Omega_{\rm v0}, \Omega_{\rm dr0},\Omega_{\rm c0}, \ga_{+}, \ga_{-}, \ga_{\rm r})=(0.7499, 0.00006,  0.1999, 1.3000, 1.0099, 1.3400)$}& ($H_{0}, \ga_{s})= (70.00^{+2.57+3.54}_{-2.04-3.05}, 0.010^{+0.007+0.055}_{-0.090-0.112})$& $0.7368$\\
{($\Omega_{\rm v0}, \Omega_{\rm dr0},  \Omega_{\rm c0},\ga_{s}, \ga_{+}, \ga_{\rm r})=(0.7499, 0.00006,  0.1999, 0.010, 1.3000,  1.3400$)}& ($H_{0}, \ga_{-})= (70.00^{+2.30+3.27}_{-2.05-3.04}, 1.009^{+0.492+0.566}_{-0.289-0.7945})$& $0.7368$\\
{$( H_{0}, \Omega_{\rm dr0}, \ga_{s}, \ga_{+}, \ga_{-}, \ga_{\rm r})=(70.00,  0.00006, 0.010, 1.3000, 1.0099, 1.3400$)}& ($\Omega_{\rm v0}, \Omega_{\rm c 0})= (0.729^{+0.167+0.218}_{-0.102-0.152}, 0.239^{+0.210+0.355}_{-0.281-0.364})$& $0.7368$\\
{($H_{0}, \Omega_{\rm v0}, \Omega_{\rm dr0}, \ga_{+}, \ga_{-}, \ga_{\rm r})=(70.00, 0.7499, 0.00006, 1.3000, 1.0099, 1.3400$)}& ($\Omega_{\rm c 0}, \ga_{s} )= ( 0.239^{+0.215+0.267}_{-0.091-0.115}, 0.010^{+0.225+0.486}_{-0.170-0.271})$& $0.7368$\\
{($H_{0}, \Omega_{\rm c0}, \ga_{s}, \ga_{+}, \ga_{-}, \ga_{\rm r})=(70.00,0.1999,  0.010, 1.3000, 1.0099, 1.3400$)}& ($\Omega_{\rm v 0}, \Omega_{\rm dr 0})= (0.729^{+0.140+0.182}_{-0.076-0.120},  0.00006^{+0.08098+0.20684}_{-0.06068-0.68741})$& $0.7368$\\
{($H_{0}, \Omega_{\rm c0}, \Omega_{\rm dr0}, \ga_{+}, \ga_{-}, \ga_{\rm r})=(70.00, 0.1999, 0.00006, 1.3000, 1.0099, 1.3400$)}& ($\Omega_{\rm v 0}, \ga_{s}  )= (0.729^{+0.076+0.094}_{-0.030-0.051}, 0.010^{+0.126+0.163}_{-0.223-0.186})$& $0.7368$\\
{($H_{0},\Omega_{\rm v0},\Omega_{\rm c0}, \ga_{+}, \ga_{-}, \ga_{\rm r})=(70.00, 0.7499, 0.1999, 1.3000, 1.0099, 1.3400$)}& ($\Omega_{\rm dr 0}, \ga_{s} )= (0.00066^{+0.03791+0.06316}_{-0.08401-0.10266},  0.010^{+0.271+0.335}_{-0.189-0.303})$& $0.7368$\\
{($H_{0},\Omega_{\rm c0}, \Omega_{\rm dr0},\ga_{s}, \ga_{-}, \ga_{\rm r})=(70.00,  0.1999,0.00006 , 1.3000, 1.0099, 1.3400$)}& ($\Omega_{\rm v 0}, \ga_{+})= (0.729^{+0.054+0.065}_{-0.08-0.014}, 1.30^{+0.147+2.60}_{-0.147-2.60})$& $0.7368$\\
{($H_{0},\Omega_{\rm c0}, \Omega_{\rm dr0},\ga_{s}, \ga_{+}, \ga_{\rm r})=(70.00,  0.1999,0.00006 , 0.010, 1.3000, 1.3400$)}& ($\Omega_{\rm v 0}, \ga_{-})= (0.729^{+0.061+0.070}_{-0.023-0.039}, 1.009^{+0.532+0.646}_{-0.532-0.646})$& $0.7368$\\
\hline
\end{tabular}}
\caption{\scriptsize{ Observational bounds for the 2D C.L. obtained in  Fig. (I) by varying two cosmological parameters. The $\chi^{2}_{\rm d.o.f}$ in all the cases studied is less than unity.}}
\label{I}
\end{minipage}
\end{table}
\end{center}

We start our statistical estimations by performing a global analysis on the eight parameters that characterize the model. In doing so, we find  that $\chi^{2}(\theta)$ reaches a minimum at  $\{H_{0},\ga_{\rm s},\ga_{+},\ga_{-},\ga_{\rm r},\Omega_{\rm  v0},\Omega_{\rm dr0},\Omega_{\rm c0}\}=(70.00, 0.010, 1.300, 1.009, 1.3400, 0.749, 0.00006, 0.199  )$ along with  $\chi^{2}_{\rm d.o.f}=20.172/(29-8)\simeq 0.9605 <1$. The two-dimensional C.L. obtained with the standard $\chi^{2}$ function
for two independent parameters is shown in Fig. (I), whereas the estimation
of these cosmic parameters is briefly summarized in Table (\ref{I}). We see that
$\ga_{s}$ varies from $ 0.010^{+0.007+0.055}_{-0.090-0.112}$ to $ 0.010^{+0.271+0.335}_{-0.189-0.303}$ at $1\sigma-2\sigma$ confidence levels, so  these values clearly fulfill
the constraint $\ga_{s}<2/3$ at $95\%$ C.L. that ensure the existence of accelerated phase of
the universe at late times [Table (\ref{I})]. Regarding the latter results, it must be
stressed that we report for the most relevant minimization
procedures the corresponding marginal  $1\sigma--2\sigma$ error bars \cite{Bayes},
as can be seen in  Table (\ref{I}). We find the best fit at $ (H_{0},\Omega_{\rm v0})= (70.00^{+4.99+6.07}_{-2.03-3.16} {\rm km~s^{-1}\,Mpc^{-1}}, 0.7299^{+0.057+0.0658}_{-0.0109-0.0238})$ with $\chi^{2}_{\rm d.o.f}=0.736$
by using the priors $(\Omega_{\rm c0}, \Omega_{\rm dr0},\ga_{s},\ga_{+},\ga_{-},\ga_{\rm r})=(0.1999, 6\times 10^{-5}, 0.010, 1.3000, 1.0099, 1.3400)$. These findings show, in broad terms,  
that the  estimated values of $H_{0}$ and $\Omega_{\rm  v0}$ are in agreement with 
the standard ones reported by the WMAP-7 project \cite{WMAP7}. The value of $\Omega_{\rm v 0}$
is slightly greater than the standard one of $0.7$ being such discrepancy less or equal to  $0.02\%$.
Moreover, we find that using the priors $( H_{0},\Omega_{\rm dr0},\ga_{s},\ga_{+},\ga_{-},\ga_{\rm r})=(70.00 {\rm km~s^{-1}\,Mpc^{-1}}, 6\times 10^{-5}, 0.010, 1.3000, 1.0099, 1.3400)$
the best-fit values for the present-day density parameters are considerably improved, namely, these turn give  
($\Omega_{\rm v0}, \Omega_{\rm c 0})= (0.729^{+0.167+0.218}_{-0.102-0.152}, 0.239^{+0.210+0.355}_{-0.281-0.364})$ along with the same goodness condition ($\chi^{2}_{\rm d.o.f}=0.7368$) [Table (\ref{I})].  Regarding the estimated value of $\Omega_{\rm c0}$, we find that it varies from
$0.239^{+0.086+0.101}_{-0.019-0.042}$ to $0.239^{+0.215+0.267}_{-0.091-0.115}$ at $68\%$, $95\%$ C.L., without showing a significant difference with the standard ones \cite{WMAP7}. In 
performing the statistical analysis, we find that $H_{0} \in [70.00^{+2.57+3.54}_{-2.04-3.05}, 70.00^{+4.99+6.07}_{-2.03-3.16}]{\rm km~s^{-1}\,Mpc^{-1}}$ 
so the estimated values are met  within $1 \sigma$ C.L.  reported
by Riess \emph{et al} \cite{H0}, to wit,  $H_{0}=(72.2 \pm 3.6){\rm km~s^{-1}\,Mpc^{-1}}$.  These values are consistent with the analysis of ACT and WMAP-7 data that 
gives $H_0 = 69.7 \pm 2.5{\rm km~s^{-1}\,Mpc^{-1}}$ \cite{do} or with  the median statistic  $H_{0}= 68\pm 2.8 {\rm km~s^{-1}\,Mpc^{-1}}$ reported in \cite{H02}.

For the sake of completeness,
we also report bounds for other cosmological relevant parameters [see Table (\ref{II})], such as the fraction of dark matter $\Omega_{\rm c}(z=0)$, the overall  equation of state at $z=0$ ($\omega_{\rm ove0}=\ga_{\rm ove0}-1$), decelerating parameter  at the present time $q_{0}$, and the transition redshift ($z_{\rm t}$) among many others, all these quantities are derived
using the ten best-fit values reported in Table (\ref{I}). We find that the transition redshift is of the order unity   $z_{\rm t}=0.604^{+0.203+0.325}_{-0.025-0.043}$ at $1\sigma, 2\sigma$ C.L., such values are close to $z_{\rm t}=0.69^{+0.20}_{-0.13}$ reported in \cite{Zt1}, \cite{Ztn}  quite recently or with the marginalized best-fit values $z_{\rm t}=0.623^{+0.039
}_{−0.052}$ listed in \cite{mizt}, \cite{mizt2}. Moreover, taking into account
a $\chi^{2}$-statistical analysis made in the $(\omega_{\rm 0}, z_{t})$--plane based on the supernova sample 
(Union 2) it has been  shown that at  $2 \sigma$ C.L.  
the transition redshift  varies from $0.60$ to $1.18$ \cite{Zt2}.
The  behavior of decelerating parameter with redshift is shown in Fig. (II), in particular, its present-day
value varies as  $q_{0}=-0.579^{+0.350+0.470}_{-0.247-0.354}$ at $68\%$, $95\%$ C.L. for the 12-cases mentioned in Table (\ref{I}), all these values are in perfectly
agreement with the one reported by WMAP-7 project \cite{WMAP7}. The aforesaid value  is similar to $q_{0} = -0.53^{+0.17}_{-0.13}$  estimated in \cite{Waga} or with the marginalized best-fit values $q_{0}=-0.671^{+0.120}_{-0.283}$ listed in \cite{mizt}, \cite{mizt2}.

The  effective equation of state (EOS) for the three components are obtained from  Eqs.(\ref{04})-(\ref{06}), they read 
\be
\label{re4}
\omega_{\rm effj}=\Big(\ga_{\rm j}-\frac{Q_{\rm j}}{\ro_{\rm j}} \Big)-1,
\ee
where $j=\rm \{x,c,dr\}$. Besides, the overall EOS of the mix and the EOS  of the scalar field are given by 
\[\omega_{\rm ove}=\frac{\ga_{\rm v}\Omega_{\rm x}+\ga_{\rm m}\Omega_{\rm c}+\ga_{\rm sm}\Omega_{\rm dr}+\ga_{\rm r}\Omega_{\rm r}}{\sum_{i}\Omega_{\rm i}}-1,  \]
\be
\label{re5} 
\omega_{\phi}=\frac{Q_{\rm m}+\ro_{\rm dr}-\ro_{\rm x}}{\ro_{\phi}}.
\ee

In Fig. (II) we plot the overall equation of state as a function of
redshift for the best-fit value shown in Table (\ref{I}),  in general,   we find that $-1\leq \omega_{\rm ove}\leq 0$, which is tantamount to saying that  $\omega_{\rm eff}(z)$ leads always to $ 0\leq (4/3)\Omega_{\rm r}+  \Omega_{\rm c}+ 2\Omega_{\rm dr} \leq 1$ while $\omega_{\phi}\geq 0$ for $z \geq0$ but it reaches negatives values in the remote future $z \in (0, -1]$, then it does not pass the values $-1$ because it fulfills the condition $Q_{\rm m}+\ro_{\rm dr}-\ro_{\rm x} \geq -1$ so these EOS do not exhibit a quintom behavior \cite{NQ}, as a matter of fact  their present-day values  are $\omega_{\rm ove0}=-0.752^{+0.137+0.345}_{-0.201-0.287}$ and  $\omega_{\phi0}=0.279^{+1.191+1.987}_{-1.552-2.015}$, respectively.  On the other hand, the effective EOS associated to the dark energy, $\omega_{\rm v}$,  does not cross the phantom divide line also and  its present-day value varies around $\omega_{\rm v0}=-2.27$  with an error  less than $0.56\%$. The present-day value of effective EOS of dark matter ($\omega_{\rm effc0}$) shows a  deviation form zero [see Fig. (II)] and  covers the range $\omega_{\rm effc0}=-3.890^{+2.378}_{-1.631}$ [see Table (\ref{II})];  these values  deviates from the cosmological constraints on the matter equation of state obtained with the five-year survey of the WMAP satellite of $\omega_{\rm effc0}={-0.35}^{+1.17}_{-0.98} \times 10^{-2}$ at $95 \%$ C.L \cite{EOSDMN1} or  with a comparable constraint found with the WMAP-5 project
but combined with galaxy clustering and supernovae data in \cite{EOSDMN2}, for example, it reported $\omega_{\rm effc0} \in [-1.5,1.13] \times 10^{-6}$. However, as  was pointed out in \cite{EOSDMN1}, a natural question  is if this value and the associated EOS can be physically acceptable. There are some models of dark matter particles produced by  a gas of interacting particle with  condensate that spontaneously breaks Lorentz 
invariance which  have a varying EOS that takes negative values \cite{EOSDMJUSTI}.


\begin{center}
\begin{table}
\begin{minipage}{1\linewidth}
\scalebox{0.4}{
\begin{tabular}{|l|l|l|l|l|l|l|l|l|l|l|l|}
\hline
\multicolumn{12}{|c|}{Bounds for cosmological parameters}\\
\hline
$z_{\rm t}$ & $q(z=0)$ & $\omega_{\rm ove}(z=0)$ &$\omega_{\phi}(z=0)$&$\omega_{\rm effv}(z=0)$&$\omega_{\rm effc}(z=0)$&$\Omega_{\rm rad}(z=0)$& $\Omega_{\rm \phi}(z \simeq 1100)$ 
&$\Omega_{\rm rad}(z \simeq 1100)$ &$\Omega_{\rm \phi}(z \simeq 10^{10})$ &$\Omega_{\rm rad}(z \simeq 10^{10})$  &$\Omega_{\rm dr}(z \simeq 10^{10})$ \\
\hline
{$0.604$}& $-0.579$& $-0.752$& $0.271$ & $-2.27$ & $-3.89$ &$0.049$ & $0.014$& $0.833$ &$0.0025$ &$0.971$ &$0.0054$\\
\hline
\end{tabular}}
\caption{\scriptsize{Derived  bounds for cosmic parameters using the best fit values of  2D C.L. obtained in  Table. (\ref{I}).  The above values represent  the mean value of each  parameter.}}
\label{II}
\end{minipage}
\end{table}
\end{center}
Regarding the behavior of  density parameters  $\Omega_{\rm x} \simeq \Omega_{\rm \phi} $, $\Omega_{\rm c}$, $\Omega_{\rm dr}$ and  $\Omega_{\rm r}$ , we find that nearly close to $z=0$ the dark energy is the main agent that speeds up the universe, far away from $z=1$
the universe is dominated by the dark matter and at very early times the radiation component enter in action, controlling the entire dynamic
of the universe around $z \simeq 10^3$[cf. Fig. (III)]. As  was expected the fraction of  dark radiation and standard radiation  at the present moment are negligible.

As is well known, distance indicators can be used for confronting distance measurements to the corresponding
model predictions. Among the most useful ones are those objects of known
intrinsic luminosity such as  standard candles, so that the corresponding comoving
distance can be determined. That way, it is possible to reconstruct the Hubble expansion rate by searching this
sort of object at different redshifts. The most important
class of such indicators is type Ia supernovae.  Then, we would like to   compare the Hubble data  with the
Union 2 compilation of 557 SNe Ia \cite{amanu} by contrasting theoretical distance modulus with the observational data set. In order to do that, we note that the apparent magnitude of a supernova placed at a given
redshift $z$ is related to the expansion history of the Universe through the distance modulus
\be
\n{mu}
\mu\equiv m -M= 5\log \frac{d_{L}(z)}{h}+\mu_{0},
\ee
where $m$ and $M$ are the apparent and absolute magnitudes, respectively, $\mu_{0}=42.38$, $h=H_{0}/100\rm{km^{-1} s^{-1}}$, and $d_{L}(z)=H_{0}(1+z)r(z)$, $r(z)$  being the comoving distance, given for a FRW metric by 
\be
\n{r}
r(z)=\int^{z}_{0}{\frac{dz'}{H(z')}}
\ee
Using the Union 2 data set, we will obtain 12 Hubble diagrams and compare each of them with the theoretical  distance modulus curves  that represent the best-fit cosmological models found with the updated Hubble data (see Fig. (IV)); it turned out that at low redshift ($z<1.4$) there is excellent  agreement between the theoretical model and the observational data.

Now, we seek for another kind of constraint that comes form the physics at early times because this can be considered as a complementary tool for testing
our model. As is well known the fraction of dark 
energy at recombination epoch should fulfill the bound $\Omega_{\rm ede}(z\simeq 1100)<0.1$ in order to the dark energy model be consistent 
with the big-bang nucleosynthesis (BBN) data.   Some signals could arise from the early dark energy (EDE) models uncovering the nature of DE as well as 
their properties to high redshift, giving an invaluable guide to the physics behind the recent speed up of the universe \cite{EDE1}. Then,  it was examined the current and future data  for constraining the amount  of EDE, the cosmological data analyzed has led to an upper bound of  $\Omega_{\rm ede}(z\simeq 1100)<0.043$ with $95\%$ confidence level (CL) in case of relativistic EDE while for a quintessence type of EDE has given $\Omega_{\rm ede}(z\simeq 1100)<0.024$ although   the EDE component is not preferred, it is also not 
excluded from the current data \cite{EDE1}. Another forecast for  the bounds of the EDE  are obtained with the  Planck  and CMBPol experiments\cite{EDE2}, thus  assuming a $\Omega_{\rm ede}(a \simeq 10^{-3}) \simeq 0.03$  for studying 
the stability of this value, it found that $1\sigma$ error coming from Planck experiment 
is $\sigma^{\rm Planck}_{\rm ede} \simeq 0.004$ whereas the CMBPol  improved this bound by a factor 4 \cite{EDE2}.  Taking into account the best-fit values reported in Table (\ref{I}), we find that at  early times the dark energy  changes rapidly with the redshift $z$ over the interval $[10^{3}, 10^{10}]$; indeed Table (\ref{II})  shows that  around $z \simeq 1100$ (recombination)  $\Omega_{\rm \phi} \simeq 0.014$. Let us  compare our estimations with the recent bounds on early dark energy  from the cosmic microwave background  using data from the WMAP satellite on large angular scales and South Pole Telescope  on small angular scales reported by  Reichardt \emph{et al} in \cite {Reic}.  They found a strong upper limit on the EDE density of $\Omega_{\rm ede} < 0.018$ at $95\%$ confidence, a factor of three improvement over WMAP data alone  \cite {Reic}. Interestingly enough, our findings  are in agreement with the aforesaid upper limit indicating that the amount of early dark energy cannot be greater than $1.4\%$ of the total energy density in the recombination epoch.  Pettorino \emph{et al}. investigated constraints on early dark energy from the CMB anisotropy, 
taking into account data from WMAP9 combined with latest small-scale measurements from the SPT. For  a simple parametrization of the time evolution of dark energy involving only two parameters, namely the fraction of dark energy at present, and the constant fraction of dark energy at early times, $\Omega_{\rm ede}$, they found a constraint $\Omega_{\rm ede} < 0.015$ at $95\%$ confidence level \cite{Amendola}. Despite our parametrization of dark energy at early times is completely different form the one proposed in \cite{Amendola}, our estimations  $\Omega_{\rm \phi} \simeq 0.014< 0.015$ meet within their upper bound.  Because early dark energy models lead to
much larger signatures in the CMB anisotropy than traditional dark energy models it is crucial to find further bounds to confront our model, where dark energy coupled to dark radiation and dark matter, with additional bounds for dark energy in the recombination era. For instance, the small-scale CMB temperature anisotropy power
measurement from the SPT bandpowers improves the constraints on the early dark energy density over
WMAP7 alone by a factor of 3.5; the $95\%$ upper limit on $\Omega_{\rm ede}$ is reduced from $0.052$ for WMAP7-only to $0.013$ for
WMAP7+SPT. This is a $38\%$ improvement on the upper limit of $\Omega_{\rm ede} < 0.018$ reported for WMAP7+K11 \cite{Reic}. Adding low-redshift geometrical
measurements does not help constrain early dark energy, although, these data have a dramatic effect on the quality of the constraints on the late-time dark energy density
and equation of state. The upper limit is essentially unchanged at $\Omega_{\rm ede} < 0.014$ for WMAP7+SPT+BAO+SNe. The $\Omega_{\rm ede} < 0.013$ bound from WMAP+SPT is the best
published constraint from the CMB (see \cite{Hou} and reference therein). Our findings point out that the model constructed here not only fulfill the severe bound
of $\Omega_{\rm \phi}(z\simeq 1100)<0.018$ obtained from the measurements of CMB anisotropy from ACT and SPT \cite{Reic}, \cite{Amendola}, \cite{Hou}  but also is consistent with the future constraints
achievable by Planck and CMBPol experiments \cite{EDE2} as well, corroborating that the value of the cosmological parameters selected before, 
through the statistical analysis made with Hubble data, are consistent with BBN constraints. Besides,  regarding the values reached by dark energy  around $z=10^{10}$ (BBN), we find that 
 $\Omega_{\rm \phi}=0.0025$ at $1\sigma$level indicating the conventional BBN processes that occurred at temperature of $1 {\rm  Mev}$ is not spoiled because the severe
bound reported  for early dark energy $\Omega_{\rm \phi}(z\simeq 10^{10})<0.21$\cite{Cyburt} or a strong upper limit $\Omega_{\rm \phi}(z\simeq 10^{10})<0.04$ \cite{Cyburt2} are fulfilled at BBN. 

We now analyze the main differences between two cosmological scenarios where the radiation term is involved. On the one hand, we can consider a  standard model where the radiation component is taken as a free evolving component which is decoupled from dark sector, but on the other hand,  the radiation component could interact with 
the dark sector; in particular,  the latter case  is an appealing manner for extracting  a physical insight about the role played by the interaction. 
At present, dark energy dominates the whole dynamics of the universe and 
there is an obvious decoupling with radiation practically. 
 However, from a theoretical point of view,  it is reasonable to expect  that dark components can interact with other fluids of the universe substantially  in the very beginning of its evolution due to process occurred in the early universe. For instance, dark energy interacting with neutrinos was investigated in \cite{GK}.
The framework of many interacting components could provide a more  natural  arena for studying  the stringent bounds of dark energy at recombination epoch.  There could be a signal in favor of  having dark matter exchanging energy with dark energy while  radiation is treated as a decoupled component \cite{hmi1}, \cite{hmi2} or the case where  dark matter, dark energy, and radiation exchange energy. More precisely, when the universe is filled with interacting dark sector plus a decoupled radiation term, it was found that $\Omega_{\rm x}(z\simeq 1100)=0.01$ \cite{hmi1} or $\Omega_{\rm x}(z\simeq 1100)=10^{-8}$ \cite{hmi2} but if radiation is coupled to the dark sector, the amount of dark energy is drastically reduced, giving $\Omega_{\rm x}(z\simeq 1100)\simeq {\cal{O}}(10^{-11})$ \cite{jefe2}. In our model, we have found that  the amount of early dark energy  leads to  $\Omega_{\rm \phi}(z\simeq 1100)=0.014$, so the behavior of dark energy  at recombination is  considerably  much  more smooth than in the aforesaid cases \cite{hmi1}, \cite{hmi2}, \cite{jefe2}, \cite{mizt2}.

So far we have shown  significant observational  evidences supporting  the model of three interacting components plus a decoupled one. However,  we will perform a further  comparative analysis for taking into account several data sets along with  theirs  constraints on the most relavant parameters \cite{Planck2013}, \cite{WMAP9}, namely, the values  or bounds reported for  $H_{0}$, $\Omega_{\rm v0}$, $\Omega_{\rm c0}$, and $\Omega_{\rm ede}$.  

Our starting point is the present-day value of the Hubble parameter $H_{0}$.
The  Planck power spectra leads to a low value of the Hubble constant, which is
tightly constrained by CMB data alone within the $\Lambda$CDM model. From the Planck+WP+highL analysis, it found that $H_{0}   = (67.3\pm 1.2) {\rm km s^{-1} Mpc^{-1}}$ at $68\%$ \cite{Planck2013}. A low value of $H_{0}$   has been found in other CMB experiments, most notably from the recent WMAP-9 analysis. Fitting
the base $\Lambda$CDM model for the WMAP-9 data, it is found $ H_{0} =(70.0 \pm  2.2) {\rm km s^{-1} Mpc^{-1}} $ at $68\%$ C.L. \cite{WMAP9}. Then, our best estimation $ H_{0} =70.00^{+3.76}_{-2.02} {\rm km s^{-1} Mpc^{-1}} $ at $68\%$ C.L  is perfectly in agreement with the value reported by WMAP-9 project but  shows a slightly difference with the  Planck+WP+highL data, which is less than $0.04\%$. Besides, Riess \emph{et al} have used HST observations
of Cepheid variables in the host galaxies of eight SNe Ia to calibrate the supernova magnitude-redshift relation\cite{Riess2011}. Their “best
estimate” of the Hubble constant, from fitting the calibrated SNe
magnitude-redshift relation, is $ H_{0}    =  (73.8 \pm 2.4) ) {\rm km s^{-1} Mpc^{-1}}$, where the  $1\sigma$ error includes known sources of systematic
errors. Therefore, this measurement is discrepant with the Planck estimate at about the 2.5$\sigma$ level. On the other
hand,  the  Spitzer Space Telescope mid-infrared observations helped  to  recalibrate secondary distance methods  used in the HST Key
Project, which led to  $H_{0}   = [74.3 \pm 1.5 ({\rm statistical}) \pm  2.1 ({\rm systematic})] {\rm km s^{-1} Mpc^{-1}} $. The error analysis in the latter result does not
include a number of known sources of systematic error and is very likely an underestimate \cite{Fredman}. In Fig. (V), we show other observational estimations of the Hubble constant that includes the megamaser-based distance to NGC4258, parallax measurements for 10 Milky Way Cepheids, Cepheid observations, a revised distance to the Large
Magellanic Cloud, and three estimates of $H_0$ based on “geometrical” methods corresponding to the points called UGC 3789, RXJ1131-1231, and SZ clusters (see \cite{Planck2013} and references therein).

The  approximate constraints  on the present-day value of dark matter with  $68\%$ errors show that  Wiggle-Z data give $\Omega_{\rm c0}=0.309^{+0.041}_{-0.035}$,  while Boss experiment seems to increase  the dark matter amount in $0.019\%$, thus $\Omega_{\rm c0}=0.315^{+0.015}_{-0.015}$; whereas the joint statistical analysis with the data 6dF+SDSS+ BOSS+ Wiggle-Z  leads to $\Omega_{\rm c0}=0.307^{+0.010}_{-0.011}$ at $68\%$ confidence level \cite{Planck2013}, showing that there is a discrepancy with our  estimation of dark matter, $\Omega_{\rm c0}=0.239^{+0.086}_{-0.019}$,  no bigger than  $0.24\%$  [see Fig. (VI)].  Concerning the estimations on the fraction of dark energy,  Planck+WP data indicate that the vacuum energy amount is $0.685 ^{+0.018}_{-0.016}$ at $68\%$ C.L,  Planck+WP+highL data lead to  $0.6830 ^{+0.017}_{-0.016}$ at $68\%$ C.L  whereas the joint statistical analysis on  Planck+WP+highL+ BAO  gives   $0.692\pm 0.010$ at $1\sigma$ level \cite{Planck2013}. In our case, the contour plot in the plane $\Omega_{\rm 0}-H_{0}$ leads to  $\Omega_{\rm v0}=0.7299^{+0.057}_{-0.0109}$, then the relative difference between Hubble data and Planck+WP+highL is $0.068\%$; moreover,  in the case of Planck+WP+highL+ BAO data such disagreement is reduced without exceeding  $0.054\%$ at $1\sigma$ level [see Fig. (VII)].  Besides, the CMB anisotropies measurements put further  constraints on the behavior of dynamical dark energy in the recombination epoch, in particular, the latest constraints on early dark energy come from  Planck+WP+highL data and leads to $\Omega_{\rm ede} < 0.009$  at $95\%$ C.L \cite{Planck2013}. We have found that $\Omega_{\rm \phi}(z\simeq 10^{3})\simeq 0.014$, so the relative error between both estimations is  $|\Omega^{\rm Us}_{\rm \phi}-\Omega^{\rm Planck}_{\rm ede}|/|\Omega^{\rm Planck}_{\rm ede}| \simeq  0.5 \%$. The latter disagreement can be reduced in a future by including additional measurements of Hubble data points or   other data sets, allowing to improve the statistical analysis performed here.

\section{ summary and conclusions}

We have examined  a universe filled  with  interacting dark radiation,  dark matter, dark energy plus a decoupled radiation term for a spatially flat FRW spacetime. Following \cite{jefe2},  we have coupled those components with a linear transversal interaction because it gives a unique  preferred direction in the interacting constraint plane  $\sum_{i}{Q_{i}}=0$ and obtained their energy densities in terms of the scale factor. We have asymptotically reconstructed the behavior of the scalar field and potential energy from the early to late eras. 

On the observational side, we have  examined the previous model  by constraining the cosmological parameters with the Hubble data and the well-known bounds  for dark energy at recombination era.  In the case of two-dimensional (2D) C.L., we have made ten statistical constraints with the Hubble function [see Fig. (I) and Table (\ref{I})]. We have found that $\ga_{s}$ varies from from $ 0.010^{+0.007+0.055}_{-0.090-0.112}$ to $ 0.010^{+0.271+0.335}_{-0.189-0.303}$ within the marginalized $1\sigma-2\sigma$ confidence levels, so  these values clearly fulfill
the constraint $\ga_{s}<2/3$ for getting an accelerated phase of the universe at late times. Using  $( H_{0},\Omega_{\rm dr0},\ga_{s},\ga_{+},\ga_{-},\ga_{\rm r})=(70.00 {\rm km~s^{-1}\,Mpc^{-1}}, 6\times 10^{-5}, 0.010, 1.3000, 1.0099, 1.3400)$ the best-fit values for the present-day density parameters  are given by ($\Omega_{\rm v0}, \Omega_{\rm c 0})= (0.729^{+0.167+0.218}_{-0.102-0.152}, 0.239^{+0.210+0.355}_{-0.281-0.364})$ along with $\chi^{2}_{\rm d.o.f}=0.7368$ [Table (\ref{I})]. We have obtained that estimated value of $\Omega_{\rm c0}$ varies from
$0.239^{+0.086+0.101}_{-0.019-0.042}$ to $0.239^{+0.215+0.267}_{-0.091-0.115}$ at $68\%$, $95\%$ C.L., without showing a significant difference with the standard ones \cite{WMAP7}. Besides, it turned  out that  $H_{0} \in [70.00^{+2.57+3.54}_{-2.04-3.05}, 70.00^{+4.99+6.07}_{-2.03-3.16}]{\rm km~s^{-1}\,Mpc^{-1}}$   so the latter values are met  within $1 \sigma$ C.L.  reported by Riess \emph{et al} \cite{H0}, are consistent with the analysis of ACT and WMAP-7 data that give 
 $H_0 = 69.7 \pm 2.5{\rm km~s^{-1}\,Mpc^{-1}}$ \cite{do} and with  the median statistic  $H_{0}= 68\pm 2.8 {\rm km~s^{-1}\,Mpc^{-1}}$ reported in \cite{H02}. Regarding the derived cosmological parameters,  for instance, the transition redshift turned out  of the order unity varying over the interval $z_{\rm t}=0.604^{+0.203+0.325}_{-0.025-0.043}$ at $1\sigma, 2\sigma$ C.L., such values  are in agreement with $z_{\rm t}=0.69^{+0.20}_{-0.13}$ reported in \cite{Zt1}-\cite{Ztn}, meet  within the $2 \sigma$ C.L  obtained with the supernovae (Union 2) data in \cite{Zt2}, and meet with the marginalized best-fit values $z_{\rm t}=0.623^{+0.039}_{−0.052}$ listed in \cite{mizt}, \cite{mizt2}. Besides, the decelerating parameters $q(z=0)=-0.579^{+0.350+0.470}_{-0.247-0.354}$ at $68\%$, $95\%$ C.L. for the 12 cases mentioned in Table (\ref{I}), all these values are in perfect
agreement with the one reported by WMAP-7 project \cite{WMAP7}. The aforesaid value  is similar to $q_{0} = -0.53^{+0.17}_{-0.13}$  estimated in \cite{Waga} or with the marginalized best-fit values $q_{0}=-0.671^{+0.120}_{-0.283}$ listed in \cite{mizt}, \cite{mizt2}.  Concerning  the overall equation of state,  we have found  that $-1\leq \omega_{\rm ove} \leq 0$ with $\omega_{\rm ove0}=-0.752^{+0.137+0.345}_{-0.201-0.287}$  whereas and  the scalar field equation of states  $\omega_{\phi}\geq 0$  if $z \geq0$ but it reaches negatives values in the remote future $z \in (0, -1]$  and its present-day value is given by   $\omega_{\phi0}=0.279^{+1.191+1.987}_{-1.552-2.015}$ [see Table (\ref{II}) and  Fig. (II)].
On the other hand,  the present-day value of effective EOS for non-baryonic dark matter varies in the interval $\omega_{\rm effc0}=-3.890^{+2.378}_{-1.631}$ [see Table (\ref{II})];such values  deviates from the cosmological constraints obtained form  WMAP-5 project $\omega_{\rm effc0}={-0.35}^{+1.17}_{-0.98} \times 10^{-2}$ at $95 \%$ C.L \cite{EOSDMN1} or with the value $\omega_{\rm effc0} \in [-1.5,1.13] \times 10^{-6}$ obtained from the WMAP-5 project along with  a combined analysis with galaxy clustering and supernovae data reported in \cite{EOSDMN2}. However, some  physically acceptable models of dark matter particles produced by  a gas of interacting particle with  condensate  have a varying EOS that take negative values \cite{EOSDMJUSTI} [Fig. (II)].  Besides, we have found that  the fraction of dark radiation  at present moment is $\Omega_{\rm dr0} \simeq {\cal O} (10^{-5})$ for the 12 cases mentioned in Table (\ref{I}).  The dark energy amount $\Omega_{\rm \phi}(z)$ governs the dynamic of the universe near $z=0$,  whereas far away from $z=1$ the universe is dominated by the fraction of non-baryonic dark matter  $\Omega_{\rm c}(z)$  and at very early times the fraction of radiation $\Omega_{\rm r}(z)$ controls the entire dynamic of the universe around $z \simeq 10^3$[cf. Fig. (III)],  giving in  the recombination era  $\Omega_{\rm rad}(z\simeq 10^{3})\simeq 0.8330$ whereas in the nucleosynthesis epoch leads to  $\Omega_{\rm rad}(z\simeq 10^{10})\simeq 0.971$.  In order to contrast the previous analysis, performed with the updated Hubble data, with another observational data set
we use the  compilation of 557 SNe Ia \cite{amanu} called Union 2 for obtaining 12 Hubble diagrams and compare each of them with the theoretical  distance modulus curves  that represent the best-fit cosmological models found with the updated Hubble data (see Fig. (IV)); it turned out that at low redshift ($z<1.4$) there is excellent  agreement between the theoretical model and the observational data.

We have  compared the behavior of early dark energy  during the recombination era or the nucleosynthesis process occurring a high redshift $z \in [10^{3}, 10^{10}]$ as a way to confirm the physical relevance of the model proposed in this article. A severe upper bound for EDE at $z \in 10^{3}$ indicated that $\Omega_{\rm ede} < 0.018$ at $95\%$ confidence  \cite {Reic}. Pettorino \emph{et al}.  constrained  early dark energy from the CMB anisotropy, 
taking into account data from WMAP9 combined with latest small-scale measurements from the SPT, giving a better upper limit  $\Omega_{\rm ede} < 0.015$ at $95\%$ confidence level \cite{Amendola}. Despite our parametrization of dark energy at early times is completely different form the one proposed in \cite{Amendola}, our estimations  lead to  $\Omega_{\rm \phi} \simeq 0.014< 0.015$, so is consistent with both  \cite {Reic} and  \cite{Amendola}. Besides, the small-scale CMB temperature anisotropy power
measurement from the SPT bandpowers improves the constraints on the early dark energy density over
WMAP7 alone by a factor of 3.5; the $95\%$ upper limit on $\Omega_{\rm ede}$ is reduced from $0.052$ for WMAP7-only to $0.013$ for
WMAP7+SPT. This is a $38\%$ improvement on the upper limit of $\Omega_{\rm ede} < 0.018$ reported for WMAP7+K11 \cite{Reic}.  Adding low-redshift geometrical
measurements does not help constrain early dark energy, although these data have a dramatic effect on the quality of the constraints on the late-time dark energy density
and equation of state. The upper limit is essentially unchanged at $\Omega_{\rm ede} < 0.014$ for WMAP7+SPT+BAO+SNe. The $\Omega_{\rm ede} < 0.013$ bound from WMAP+SPT is the best
published constraint from the CMB (see \cite{Hou} and reference therein). Our findings point out that the model constructed here not only fulfill the severe bound
of $\Omega_{\rm \phi}(z\simeq 1100)<0.018$ obtained from the measurements of CMB anisotropy from ACT and SPT \cite{Reic}, \cite{Amendola}, \cite{Hou}  but also is consistent with the future constraints
achievable by Planck and CMBPol experiments \cite{EDE2} as well, corroborating that the value of the cosmological parameters selected before, 
through the statistical analysis made with Hubble data, are consistent with BBN constraints. Regarding the values reached by dark energy  around $z=10^{10}$ (BBN), we see that 
 $\Omega_{\rm \phi}=0.0025$ at $1\sigma$level indicating the conventional BBN processes that occurred at temperature of $1 {\rm  Mev}$ is not spoiled because the severe
bound reported  for early dark energy $\Omega_{\rm \phi}(z\simeq 10^{10})<0.21$\cite{Cyburt} or a strong upper limit $\Omega_{\rm \phi}(z\simeq 10^{10})<0.04$ \cite{Cyburt2} are fulfilled at BBN.  In addition,  we have performed a full comparison between the constraints put by the Hubble data  on  $H_{0}$, $\Omega_{\rm v0}$, $\Omega_{\rm c0}$, and $\Omega_{\rm \phi}(z=1100)$ parameters with the recent bounds reported by the Planck mission and  WMAP-9 project [see Fig. (V)-Fig. (VI)-Fig. (VII)].

Finally, notice that the neutrino sector  is another fertile arena that the standard model is still unable to fully describe, with open questions related to the effective number of neutrinos. Indeed,  the presence of an extra dark radiation component could be  an indication
for an extra (sterile) neutrino species provided the cosmological constraints gave an effective number of neutrino greater than the standard value ($N_{\rm eff}>3.046$), however, this issue requires further analysis, specially taking into account the data  released by  the ATC along with the SPT \cite{Neff}.



\acknowledgments
We are grateful to the referee for his careful reading of the manuscript and for  for making useful suggestions, which  helped to improve the article.
L.P.C thanks  the University of Buenos Aires under Project No. 20020100100147 and the Consejo Nacional de Investigaciones Cient\'{\i}ficas y T\' ecnicas (CONICET) under Project PIP 114-200801-00328 for the partial support of this work during their different stages. M.G.R is partially supported by Postdoctoral Fellowship program of  CONICET. 

\end{document}